\documentclass[Afour,sagev,times,doublespace]{sagej}

\usepackage{moreverb,url}
\usepackage{array}
\usepackage{bm}

\newcommand\BibTeX{{\rmfamily B\kern-.05em \textsc{i\kern-.025em b}\kern-.08em
T\kern-.1667em\lower.7ex\hbox{E}\kern-.125emX}}

\def\volumeyear{2019}

\begin{document}

\runninghead{Chambers, Colebank, Qureshi, Clipp, Olufsen}

\title{Structural and hemodynamic properties in murine pulmonary arterial networks under hypoxia-induced pulmonary hypertension}

\author{Megan J. Chambers\affilnum{1}, Mitchel J. Colebank\affilnum{1}, M Umar Qureshi\affilnum{1,2}, Rachel Clipp\affilnum{2}, Mette S. Olufsen\affilnum{1}}

\affiliation{\affilnum{1}North Carolina State University, Raleigh NC\\
\affilnum{2}Kitware, Inc., Carborro NC}

\corrauth{Mette Olufsen, 
Department of Mathematics, 
2311 Stinson Drive, 
North Carolina State University,
Raleigh, NC 27695, USA.} 

\email{msolufse@ncsu.edu}

\begin{abstract}
Detection and monitoring of patients with pulmonary hypertension, defined as a mean blood pressure in the main pulmonary artery above 25 mmHg, requires a combination of imaging and hemodynamic measurements.  This study demonstrates how to combine imaging data from microcomputed tomography (micro-CT) images with hemodynamic pressure and flow waveforms from control and hypertensive mice. Specific attention is devoted to developing a tool that processes CT images, generating subject specific arterial networks in which 1D fluid dynamics modeling is used to predict blood pressure and flow. Each arterial network is modeled as a directed graph representing vessels along the principal pathway to ensure perfusion of all lobes. The 1D model couples these networks with structured tree boundary conditions informed by the image data. Fluid dynamics equations are solved in this network and compared to measurements of pressure in the main pulmonary artery. Analysis of micro-CT images reveals that the branching ratio is the same in the control and hypertensive animals, but that the vessel length to radius ratio is significantly lower in the hypertensive animals. Fluid dynamics predictions show that in addition to changed network geometry, vessel stiffness is higher in the hypertensive animal models than in the control models.
\end{abstract}


\maketitle

\section{Introduction}

The pulmonary vasculature forms a rapidly branching network of highly compliant vessels that, in healthy subjects, conduct blood at low pressure. Blood is ejected from the right ventricle into the main pulmonary artery (MPA) and transported to every lobe within the lung. Since the lungs are situated deep in the body, it is not possible to measure blood pressure noninvasively. However, such measurements are essential to diagnose and assess the progression of pulmonary hypertension (PH), defined as a mean pressure above 25 mmHg in the MPA~\cite{Galie15}. PH is rare, but the incidence rate is increasing~\cite{Wijeratne18}, and the disease is associated with high morbidity and mortality~\cite{Chang16}. A positive diagnosis requires an assessment of chest images and blood pressure measurements. Chest images can be obtained using radiography, ultrasound, magnetic resonance imaging (MRI), or computed tomography (CT)~\cite{Marini18,Kiely19}, and blood pressure is measured invasively using right heart catheterization (RHC)~\cite{Kiely19}. To predict outcomes of interventions and assess disease progression, these data streams should be integrated. One way to do so is by using a computational fluid dynamics model to predict blood pressure and flow in networks extracted from medical images and to compare these predictions with data. In this study, we solve a one-dimensional (1D) fluid dynamics model in pulmonary arterial networks extracted from micro-computed tomography (micro-CT) images from control and hypertensive mice and compare pressure predictions with \textit{in-vivo} pressure and flow measurements.

\subsection{Geometry of pulmonary arterial vasculature}

The morphometry of vascular networks has been the topic of numerous studies~\cite{Murray26,Singhal73,Horsfield78,Zamir78,Huang96,Olufsen99,Olufsen00, Molthen04,Davidoiu16}. In 1926, Murray proposed a power law based on an optimality principle describing how vessel radii change across bifurcations~\cite{Murray26}. Murray's law predicts optimal vessel dimensions, which minimize the total work under steady flow. In 1978, Zamir~\cite{Zamir78} derived additional optimality principles, predicting a flow-radius relationship that minimizes the pumping power and lumen volume in cardiovascular networks. Zamir's optimality principles also describe how radii change across bifurcations. He introduced two ratio laws: an asymmetry ratio relating the radii of the two daughter vessels, and an area ratio relating the combined area of the daughter vessels to the area of their parent vessel. Both Murray's and Zamir's laws were inspired by data, but were derived from theoretical principles. 

The studies by Murray and Zamir were set up to characterize whole cardiovascular networks. However, since they are derived under a steady flow assumption, they are more appropriate for describing the flow in the small vessels, where pulsatility plays a minor role. Olufsen et al.~\cite{Olufsen99,Olufsen00} conducted the first study utilizing a multi-scale approach distinguishing large and small vessels. She solved nonlinear 1D fluid dynamics equations in the large vessels and used Murray's and Zamir's optimality principles to predict pressure and flow in the small vessels. The large vessels were represented by their radius, length, and connectivity within the network, while the small vessels were represented by a self-similar, asymmetric structured tree, relating the radii and length of daughter vessels to their parent. Rather than using an exponent of 3, proposed by Murray, Olufsen used an exponent of 2.76 obtained from analysis of arterial casts~\cite{Suwa63,Singhal73,Horsfield78,Pollanen92,Huang96}. These casts were generated by injecting liquid resin into the arterial network, and the vessel dimensions were measured using calipers on the hardened resin cast. While these data provide essential geometric information, they were all obtained in a single lung, and therefore do not capture the variation between individuals~\cite{Huang96}. Moreover, the casts were fragile, some pieces of the casts broke, and there is an inherent human error in measuring the dimensions with a physical tool. As noted above, Olufsen's original structured tree was informed by the cast data, i.e., it is not subject-specific. However, as discussed by Colebank et al.~\cite{Colebank19}, inter-individual variation in vessel diameters, length, and connectivity significantly impact flow predictions, highlighting the importance of generating networks encoding subject-specific geometry.

In recent years, medical imaging technologies have emerged as valuable and efficient for obtaining high fidelity measurements of vascular geometries~\cite{Molthen04,Davidoiu16}. Image-based analyses of pulmonary vascular networks have been done in a range of species using a variety of technologies. These studies provide a detailed description of the network geometry but were not used to investigate if the geometry satisfies aforementioned optimality principles. To our knowledge, only two studies (Burrowes et al.~\cite{Burrowes05} and Clark et al.~\cite{Clark18}) have generated large pulmonary subject-specific network models. These studies used imaging data to generate large arterial and venous space-filling networks. This method ensured that supernumerary vessels (small vessels that emerge at nearly 90$^{\circ}$ angles from the large arteries) were included in the model domain. In this study, we expand on these results by using data to generate self-similar networks informed by the branching structure in healthy control and hypertensive arterial networks from mice.

\subsection{Pulmonary arterial hemodynamics}

The pulmonary arteries form a rapidly bifurcating network with more than 20 generations, depending on the species~\cite{Townsley13}. Conducting nonlinear fluid dynamics simulations in subject-specific geometric networks of this size is not feasible, as the network would include more than $2^{20}$ vessels. One way to avoid this is to represent the vasculature by an artificially generated network that encodes the subject-specific branching structure. 

Several pulmonary studies have used fluid dynamics to predict flow and pressure in the large vessels. Yang et al.~\cite{Yang19} used a 3D subject-specific model to characterize the time-averaged wall shear stress in the MPA in pediatric pulmonary hypertension, and Kheyfets et al.~\cite{Kheyfets15} used a 3D subject-specific model to compute spatially averaged wall shear stress under PH conditions. While 3D models provide high fidelity predictions, they require vast computing power making it difficult to use them for clinical analysis of large datasets. At the other end of the computational spectrum are 0D models that predict blood flow and pressure using an electrical circuit analogy. As noted in the review by Tawai et al.~\cite{Tawai11}, 0D models can predict hemodynamics at a low computational cost, but they require a plethora of parameters that are not uniquely identifiable~\cite{Marquis18}. Besides, 0D models cannot predict wave propagation. PH is associated with stiffening of large and small arteries, as well as microvascular rarefaction~\cite{Simonneau04}. These morphological changes increase wave-propagation and, eventually, the load on the heart (i.e., ventricular afterload). As shown in our previous study~\cite{Qureshi18}, wave propagation can be predicted effectively using 1D models, achieving a higher fidelity prediction than the 0D model at a computational cost that is significantly lower than 3D models.

1D fluid dynamics models have been used to predict hemodynamics (flow, pressure, and cross-sectional area) in both arterial and venous networks. This model type connects large vessels, characterized by their length and radius, in a network informed by data. At the network inlet, an inflow profile, driving the 1D model, is specified from data or computed by a heart model. At the outlet, terminal vessels are coupled to the micro-circulation via boundary conditions formulated using: a) a Windkessel model (an electrical circuit with two resistors and a capacitor); b) a lumped parameter model linking the arterial network model to a closed-loop circuit; or c) a multiscale model coupling the larger arteries to a small-vessel geometric model.
 
The most commonly used 1D models use boundary conditions of type a). Studies by Fossan et al.~\cite{Fossan18} and Epstein et al.~\cite{Epstein15} found that in the systemic circulation, flow and pressure in the large arteries can be predicted from a network including the aorta and one branch off this vessel.  Colebank et al.\cite{Colebank19a} drew similar conclusions in the pulmonary circulation, showing that the model sensitivity to boundary conditions decreases with network size. These results suggest that it is not necessary to explicitly represent all vessels in the vasculature and that it is possible to separate the network into large vessels (modeled explicitly) and small vessels (represented by boundary conditions). Another result by Colebank et al.~\cite{Colebank19} showed that changes in connectivity significantly impact flow and pressure predictions in the pulmonary circulation, emphasizing the importance of generating subject-specific models. 

In the second model type, b), the arterial network model is linked to a 0D electrical circuit representation forming a closed-loop cardiovascular model. Mynard and Smolich~\cite{Mynard15} developed the most advanced model of this type. Their model represents the large arteries and veins in 1D, while the heart and small vessels are modeled using an electrical circuit. This model is ideal for studying flow and pressure in generic subjects, but due to its complexity, it is challenging to conduct subject-specific simulations.

Lastly, models of type c) utilize a multiscale approach representing the large vessels explicitly and the small vessels by fractal trees. The advantage of this approach is that it becomes feasible to predict hemodynamics in the large vessels using the full 1D Navier-Stokes equations, while small vessel hemodynamics can be computed using a simple linearized model. Olufsen~\cite{Olufsen99} developed the first model of this type, studying hemodynamics in the systemic arteries. Spilker et al.~\cite{Spilker07} extended Olufsen's results deriving a fractal tree model for the pulmonary arteries using data collected by Huang et al.~\cite{Huang96}. Spilker et al.'s work successfully estimated pressure, flow, and impedance (magnitude and phase) in pigs, but did not verify if the data published by Huang et al.~\cite{Huang96} provide a valid representation of porcine pulmonary vasculature. 

These early studies ~\cite{Olufsen99,Olufsen00,Spilker07} introduced multiscale models including both large and small vessels, but the structured trees representing the small vessels were only used to provide an impedance boundary condition (as an alternative to the Windkessel model). Recent studies by Olufsen et al.~\cite{Olufsen12} and Qureshi et al.~\cite{Qureshi14} expanded these results developing a multiscale  model predicting dynamic flow and pressure in both large and small pulmonary arteries. In these works, the vessel geometry for the large vessels was determined from data, while the fractal network was parameterized using literature data.

An alternative approach is to solve simplified equations (e.g., Poiseuille flow) in all vessels of the pulmonary tree. This approach was used by Borrowes et al.\cite{Burrowes05} who constructed a full network of the arteries and veins from imaging data.  Clark et al.\cite{Clark18} who solved the fluid dynamics equations using a linear transmission line model. This recent study added another level to the multiscale model, coupling arteries and veins to an alveolar sheet model and enabling prediction of perfusion. However, by not separating the large and small vessels, these alternative approaches ignored inertial effects in the large vessels. 

The studies discussed above use data to guide model predictions but did not fit predictions to data. Our goal is to separate the vasculature into two parts: large vessels transporting blood to every lobe in the lung through a principal pathway, and small vessels perfusing each sub-area effectively. The large vessels will be described explicitly by their length, radius, and connectivity within the network (informed by subject-specific CT images), while the small vessels will be represented by subject-specific fractal trees, using parameters extracted from micro-CT images. The latter is of importance as it will enable us to characterize remodeling with PH, which for most PH groups~\cite{Simonneau04} starts in the small vessels, increasing vessel stiffness and decreasing the area. As the disease progresses, remodeling also affects the large vessels~\cite{Molthen04}, described by large vessel parameters.

In summary, this study presents a data-driven approach to predict hemodynamics in pulmonary arterial networks extracted from control and hypertensive mice. Geometric data are obtained from micro-CT images from excised mouse lungs~\cite{Hislop76,Vanderpool11} segmented using 3D Slicer\textregistered~\cite{Fedorov12, Kikinis14, Slicer19}, skeletonized, and used to construct directed graphs. From these graphs, we extract vessels belonging to the principal pathway and determine subject-specific fractal properties in the non-principal vessels. We conduct 1D fluid dynamics simulations predicting pressure and flow in both the large and small pulmonary arteries, and compare predictions between the control and hypertensive animals. 

\section{Materials and Methods}

This study includes three components: image analysis, network generation, and hemodynamics modeling. We briefly describe protocols used for data acquisition, followed by a detailed description of the image segmentation process, construction of directed graphs, principal pathway extraction, structured tree parameter calculation, and the 1D fluids model for predicting hemodynamics.

\subsection{Experimental protocol} \label{Sec_exp}

This study analyzes micro-CT images from male C57BL6/J mice, aged 10-12 weeks. The images were made available by Naomi Chesler, University of Wisconsin-Madison, and details of the imaging protocol can be found in~\cite{Hislop76,Vanderpool11}. We analyze images from 3 control and 3 hypertensive mice, induced by keeping the mice in an hypoxic environment (FiO$_2$ reduced by half, to $10\%$) for 10 days. The mice were euthanized by exsanguination, their lungs were extracted, and the MPA was cannulated (PE-90 tubing, 1.27 mm outer and 0.86 mm inner diameter) well above the first bifurcation. The pulmonary arteries were perfused with perfluorooctyl bromide at a pressure of 7.4 mmHg and placed in a micro-CT scanner. Lungs were rotated 360$^\circ$ in the imaging chamber, and planar images were obtained at 1$^\circ$ increments, resulting in 360 planar images. The planer images were reconstructed using the Feldkamp cone-beam algorithm and converted into 3D volumetric datasets and stored as Digital Imaging and Communications in Medicine (DICOM) 3.0 images. 

Dynamic pressure-flow data was recorded as described by Tabima et al.~\cite{Tabima12} and used for computational fluid dynamics simulations. Hemodynamic waveforms were measured in control and hypertensive adult male C57BL6/J mice aged 12-13 weeks. Hypertension was induced by keeping the animals in a hypoxic environment (FiO$_2$ reduced by half, to $10\%$) for 21 days. The pressure was measured using a 1.0-F pressure-tip catheter (Millar Instruments, Houston, TX), and the flow was measured by ultrasound (Visualsonics, Toronto, Ontario, CA). For each animal, an ensemble average waveform was computed and sampled at 30 MHz. 

\begin{figure}
\centering
 \hspace{-0.5cm}\includegraphics[width=0.9\columnwidth]{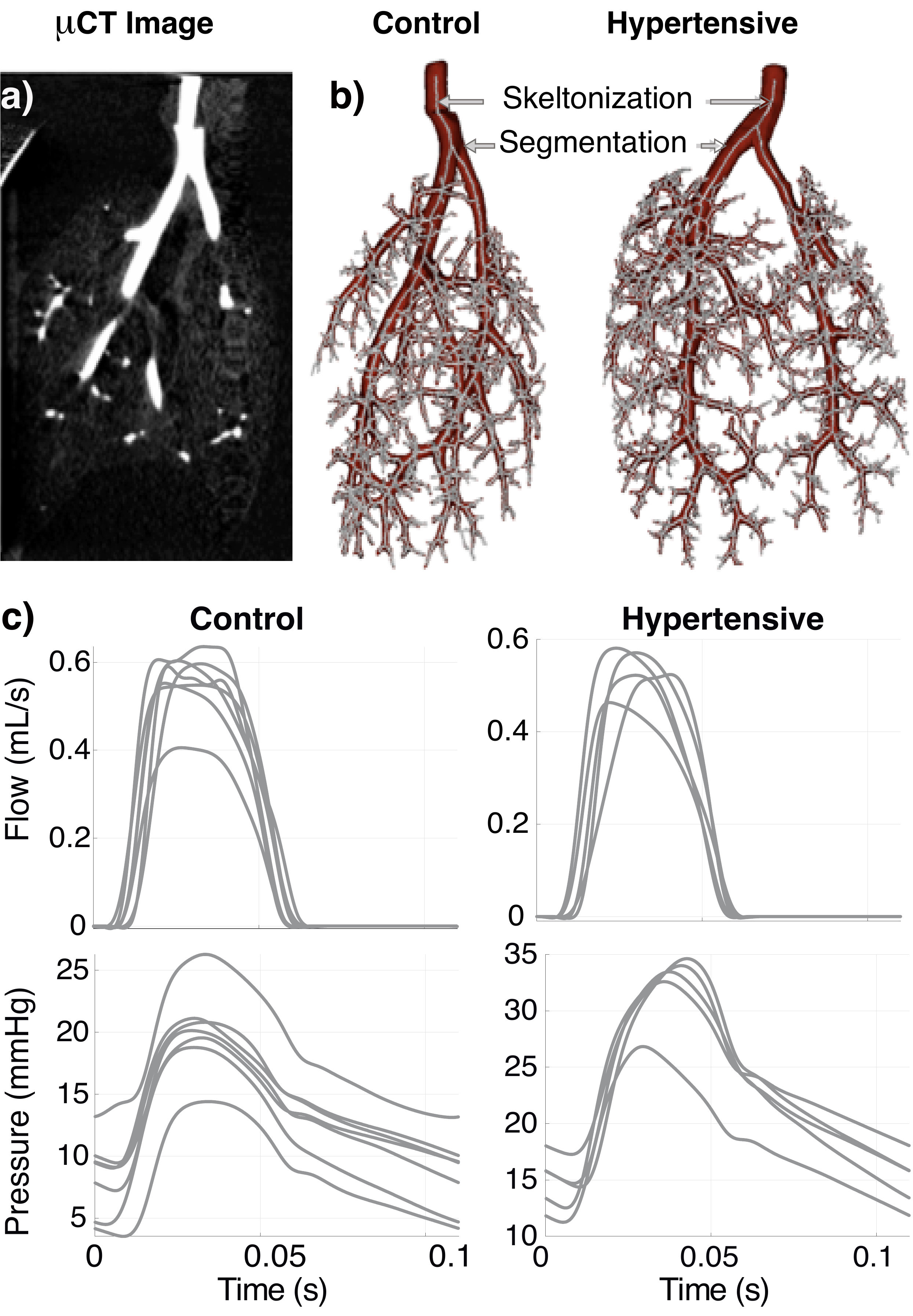}
\caption{(a) Micro-CT image from a control mouse. (b) 3D segmentation and skeletonization (one-voxel wide) networks from representative control and hypertensive mouse. (c) Main pulmonary artery flow (ml/s) and pressure (mmHg) waveforms from 7 control and 5 hypertensive animals.}
\label{fig:HemodynData}
\end{figure}

\subsection{Image segmentation}

\begin{figure}
\centering
\includegraphics[width=6cm]{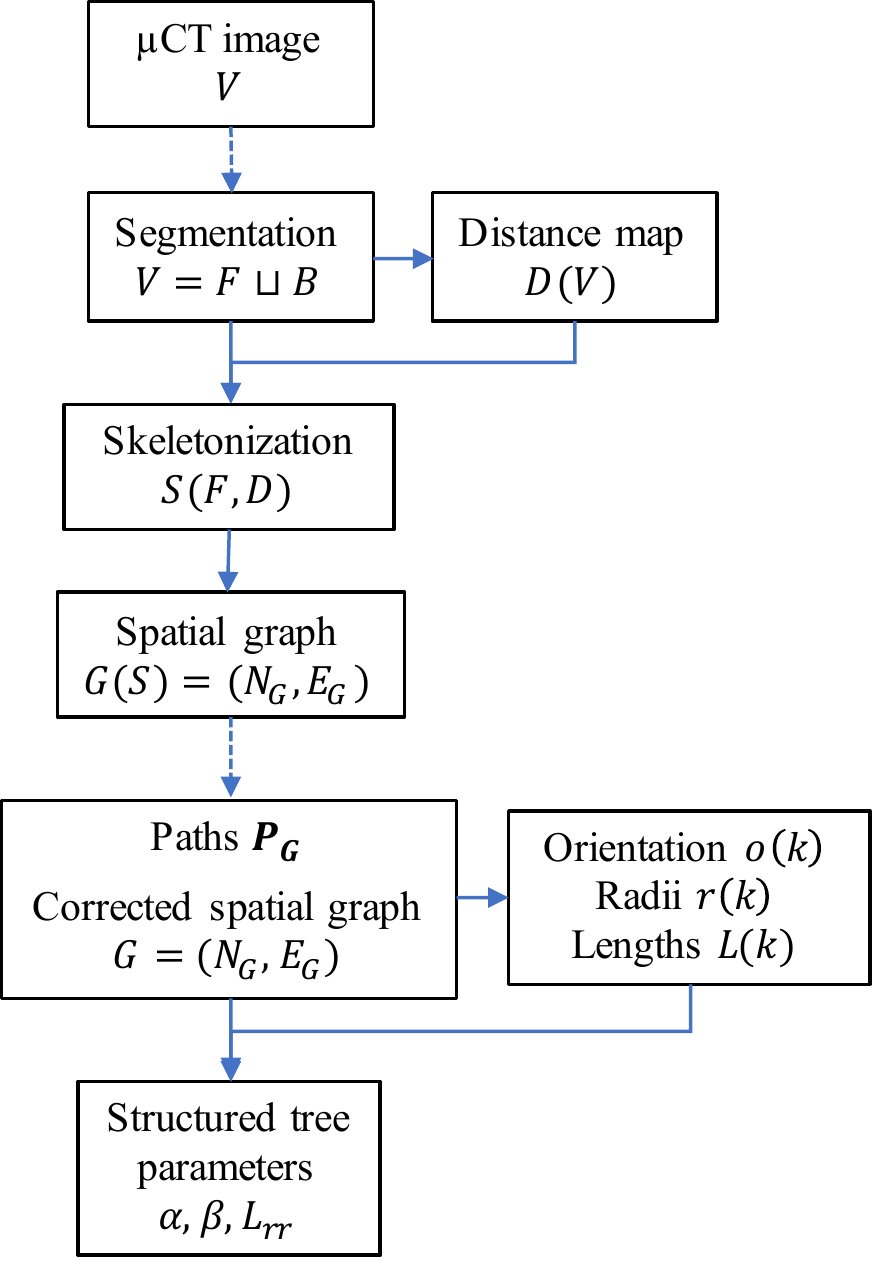}
\caption{Workflow for the segmentation and graph extraction process. Solid lines represent processes that are fully automated, and dashed lines represent processes requiring user interaction.}
\label{fig:workflow}
\end{figure}

A 3D volumetric representation (shown in Fig.~\ref{fig:HemodynData}b) was rendered from micro-CT scans (shown in Fig.~\ref{fig:HemodynData}a). To construct a vascular network encoded by a directed graph, the 3D representation was skeletonized (shown in Fig.~\ref{fig:HemodynData}b) using the image analysis process outlined in Fig.~\ref{fig:workflow}.

Each micro-CT image comprises a finite set of voxels (vx) called a \emph{voxel complex}. The set of all voxel complexes is denoted by $\mathbb{V}^3$, with the individual complexes denoted by $V\in\mathbb{V}^3$. The dimensions of each image are $497$ vx $\times$ $497$ vx $\times$ $497$ vx, and each voxel is $1$ mm $\times$ $1$ mm $\times$ $1$ mm. Each voxel $v\in V$ has spatial coordinates $(x_v,y_v,z_v)$ and intensity value $I(v)\in[0,255]$, with $0$ and $255$ denoting the intensity of a black and a white voxel, respectively.  In general, voxels representing anatomical features will have higher intensities than other voxels.

The \emph{segmentation} process involves isolation of the pulmonary arterial tree. This is done by partitioning $V$ into a set $F$ of ``foreground" voxels (belonging to the tree) and a set $B$ of ``background" voxels,  $V=F\sqcup B$. We use the open-source program 3D Slicer\textregistered~\cite{Fedorov12, Kikinis14, Slicer19} to segment and construct a 3D representation of the foreground voxels (Fig.~\ref{fig:HemodynData}b). This is done using global thresholding and manual editing. 

Global thresholding requires specification of a lower ($\tau_\ell$) and an upper ($\tau_u$) intensity. Every voxel with $I(v)\in[\tau_\ell,\tau_u]$ is included in the foreground $F$. Thresholds are selected \emph{ad hoc} to ensure the entire tree is included in $F$. Recall that the lung is excised and placed in a cylindrical hypobaric chamber before imaging, and that the arteries are perfused with a contrast, i.e., the high-intensity voxels belong to either the arterial tree or to the cylinder. As a result, we only need to specify the lower threshold $\tau_\ell$, while keeping the upper threshold at $\tau_u=255$. To isolate the arterial network, we manually remove the voxels representing the cylinder from $F$. Also, the cannula can distort the appearance of the MPA, which can mislead the skeletonization. To prevent identifying a false root vessel, manual editing was performed to smooth the MPA, without distorting the vessel geometry. 

\subsection{Skeletonization and graph extraction}

The result of the segmentation process is a 3D volumetric representation of the foreground $F$ (representing the arterial tree). The next steps needed to generate a spatial graph include obtaining 1) a distance map and 2) a centered skeletonization. Using these, we design a subject-specific directed graph (including all vessels visible in the image).

\paragraph{The distance map} $D(V)$, or simply $D$, is a voxel complex with the same spatial dimensions as $V$, which encodes distance. For all voxels in the foreground $F$, we compute the Euclidean distance between any two voxels $u,v\in V\in\mathbb{V}^3$ as
\begin{equation} 
\left\Vert u-v \right\Vert_2=\sqrt{(x_u-x_v)^2+(y_u-y_v)^2+(z_u-z_v)^2}
\label{eq:euclid}
\end{equation}
For a segmentation $V=F\sqcup B$, we define the function \\$d:V\to \mathbb{R}$ in equation (\ref{eq:dmap}).
\begin{equation} 
d(v)=\min_{u\in B}\left\Vert u-v \right\Vert_2
\label{eq:dmap}
\end{equation}
For each $v\in V$, there exists a voxel $v_D\in D$ with spatial coordinates $(x_v,y_v,z_v)$ and intensity $I(v_D)=d(v)$.

\paragraph{The skeletonization} $S(F,D)$ refer to a thinned version of the foreground voxel complex $F$, which preserves the connection and branching pattern of $F$ (Fig.~\ref{fig:HemodynData}b). We compute $S$ using Couprie and Bertrand’s ``Asymmetric Thinning" algorithm, which iteratively removes voxels from $F$ following the framework of ``critical kernels"~\cite{Couprie15, Hernandez18, DGtal18}. This creates a thinned copy of $F$, in which each branch is one voxel wide. During the ``Asymmetric Thinning" algorithm, a choice must periodically be made to chose what voxels, within a group of voxels $X\subseteq F$, should be included in $S$. To determine which voxels $\hat{v}\in X\subseteq F$ should be kept in $S$, we define the function $Select:F\to F$ given by
\begin{equation}\label{eq:select}
    Select(X)=\{\hat{v}\}\text{ such that }d(\hat{v})=\max_{v\in X} d(v)
\end{equation}
The voxel with the maximum $d(v)$ value is the most centered. By choosing this voxel, the resulting skeletonized network $S$ is centered in $F$ (Fig.~\ref{fig:HemodynData}b)~\cite{Hernandez18}. We refer to $S$ as the ``skeletonization" or the ``centerline network" of $F$.

\begin{table}
\small\sf\centering
\caption{Graph terminology}
\begin{tabular}{m{2cm} |m{5.5cm} }
\toprule \textbf{Term} & \textbf{Definition} \\
\midrule degree & The \emph{degree} of a voxel $v\in S$, denoted $\deg(v)$, is the number of voxels in $S$ adjacent to $v$. \\  
\midrule nodes & A \emph{node} is a voxel $v\in S$ with either $\deg(v)=1$ (\emph{terminal node}) or $\deg(v)>2$ (\emph{junction node}). Nodes are given numerical \emph{node IDs}, starting at 0 and denoted $nID(v)$. The \emph{root node} of a network is the terminal node at the end of the edge representing the MPA.\\ 
\midrule edge points & An \emph{edge point} is a voxel $v\in S$ with $\deg(v)=2$ that lies along an edge between two nodes. \\
\midrule edges & An \emph{edge}, denoted $E_{(a,b)}$ is a collection of consecutively adjacent voxels between two nodes $a,b\in N_G$. Any $v\in E_{(a,b)}\cap\{a,b\}^C$ is an edge point. We write the edges in the form $E_{(a,b)}=\{[a,b]:e_1,e_2,...,e_m\}\in E_G$, where $e_i$ is the $i^{th}$ edge point along $E_{(a,b)}$. An arbitrary ordering is placed on the edges so that each $E_{(a,b)}$ has a numerical ID $k$ ranging from $1$ to $|E_G|$. We call this the \emph{vessel ID} and denote it $vID(E_{(a,b)})=k$.\\
 \midrule spatial graph & A \emph{spatial graph} $G=(N_G,E_G)$ is a collection of nodes ($N_G$) connected by a collection of edges ($E_G$). For two nodes $u,v\in S$ with $nID(u)=a$ and $nID(v)=b$, we say $a,b\in N_G$ and denote an edge between them as $E_{(a,b)}\in E_G$. \\
 \midrule small cycle & A \emph{cycle} occurs in a spatial graph $G$ if for three nodes $a,b,c\in N_G$, we have edges $E_{(a,b)}, E_{(b,c)}, E_{(c,a)}\in E_G$. A \emph{small cycle} is a cycle where each of the three edges involved consist of only their nodes and no additional edge points, ie: $E_{(\text{node}1, \text{node}2)}\cap\{\text{node}1, \text{node}2\}^C$ for all $E_{(\text{node}1, \text{node}2)}$ in the cycle. These small cycles are discounted as graphical errors rather than anatomical loops since the cycle only involves 3 adjacent voxels.\\
\bottomrule
\end{tabular}
\label{terms}\end{table}

\begin{figure*}
\centering
\includegraphics[width=0.7\textwidth]{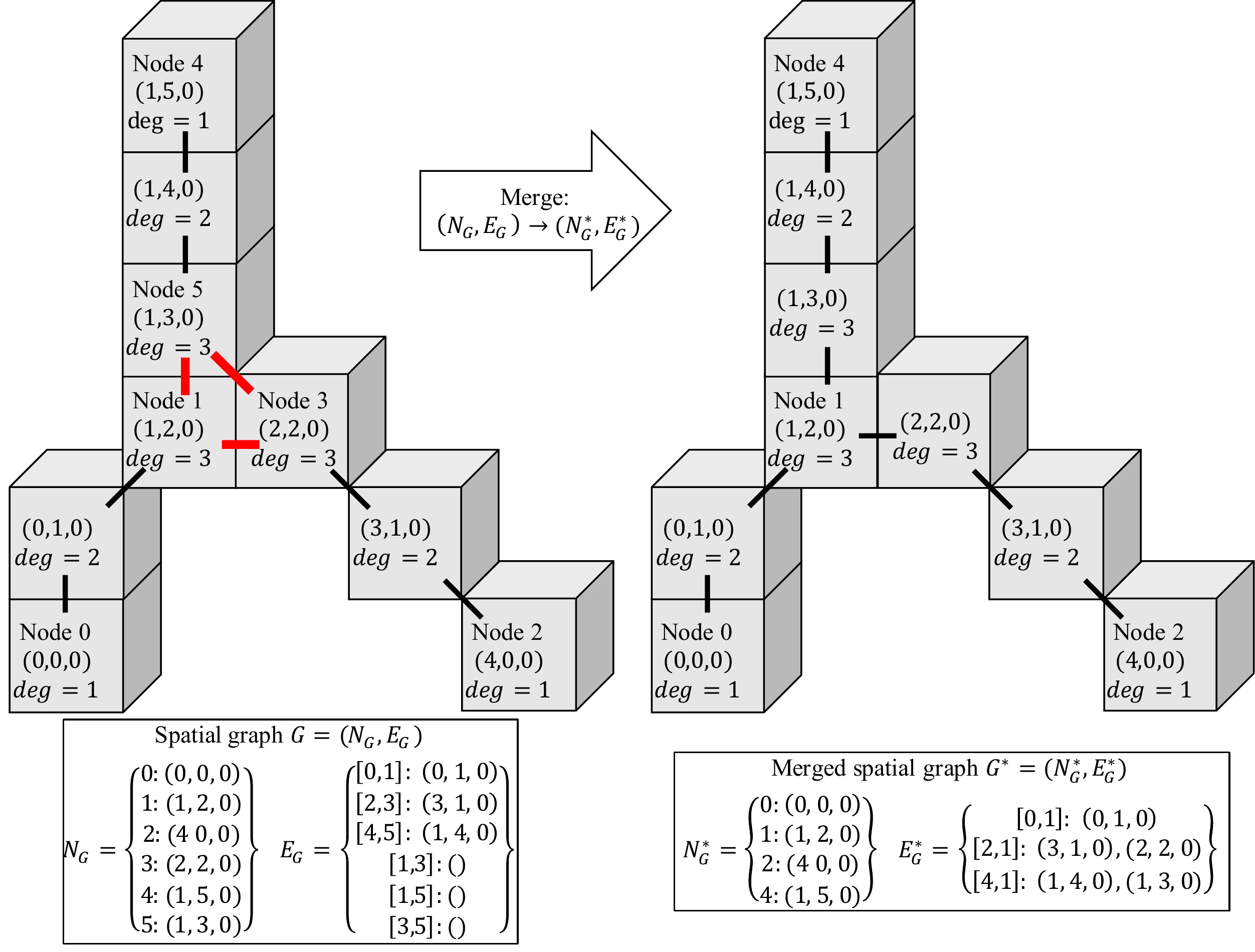}
\caption{Example spatial graph before and after merging. Voxels are labeled with coordinates, degree, and node ID. The red edges in the left graph represent a small cycle in $G$, which is broken in the merged graph $G^*$.}
\label{fig:merge}
\end{figure*}

To examine vessel branching patterns~\cite{DGtal18} we extract a \emph{spatial graph} $G(S)$, or simply $G$, from $S$.  Details of the graph extraction process are described in~\cite{Hernandez18}, and relevant graph terminology is included in Table \ref{terms}. Each voxel in $S$ corresponds to a terminal node, a junction node, or an edge point in $G$ and is connected to all adjacent voxels of $S$. Starting at one of the terminal nodes $a$, the algorithm records coordinates of adjacent edge points $e_1,e_2,...$ until another terminal or junction node $b$ is reached. Next, the nodes $a$ and $b$  are added to $N_G$ and the edge $E_{(a,b)}=\{[a,b]:e_1,e_2,...,e_m\}$ is added to $E_G$, while marking nodes that have already been visited. This is repeated for all terminal and junction nodes. To ensure that the resulting graph is a tree, small cycles of nodes of degree $\leq3$ are broken by removing the longest edge in the cycle (Fig.~\ref{fig:merge}). This process is known as ``merging". 

\subsection{Exception handling}
The skeletonization process produces errors in the graphs that do not align with the 3D representation. These errors can be categorized into four types: false branches, double edges, duplicate points, and small cycles. The false branches have to be detected manually, while the other error types can be detected automatically. 

\begin{figure*}
\centering
\includegraphics[width=0.8\textwidth]{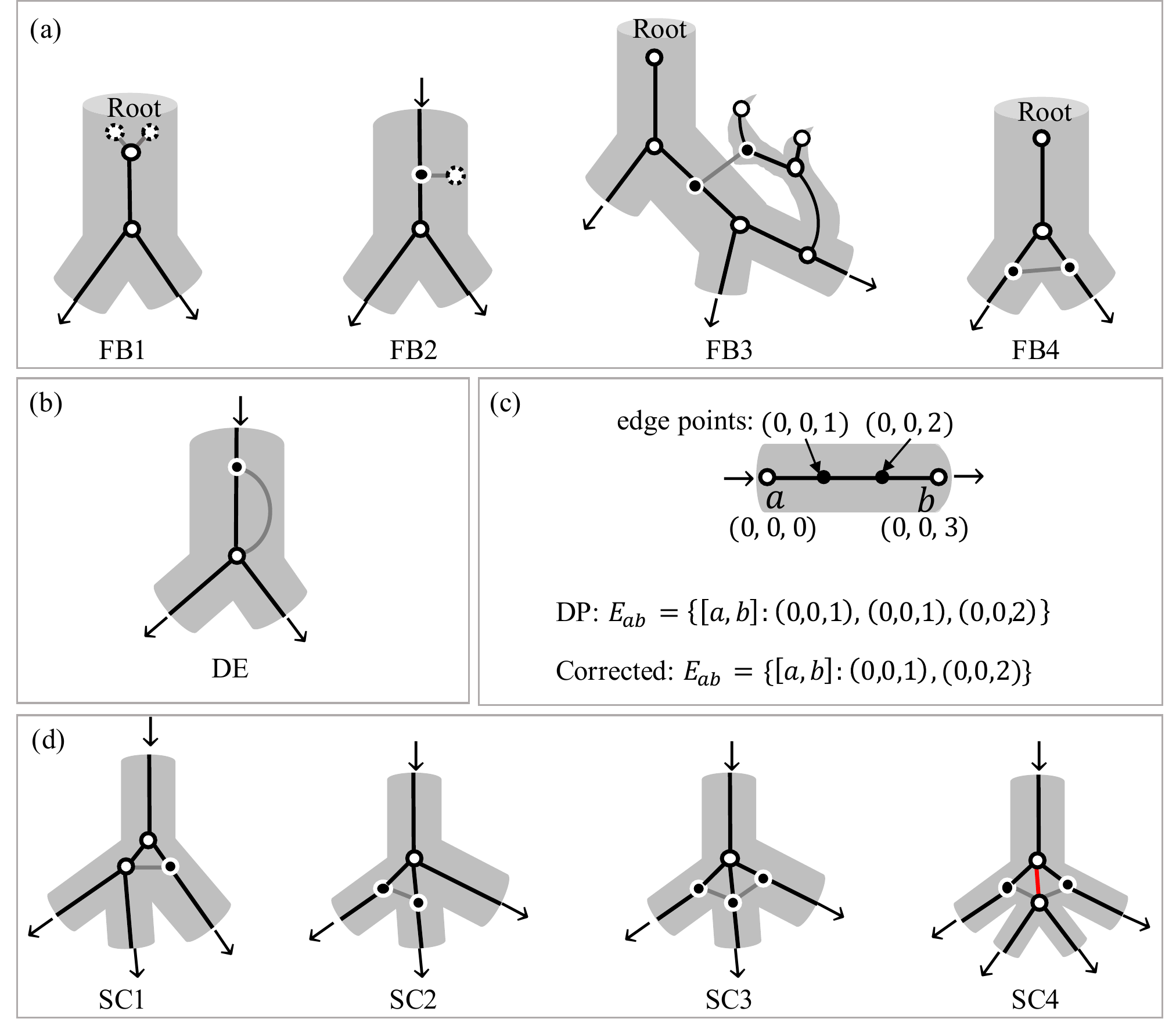}
\caption{Skeletonization gives four types of errors: (a) false branches (FB), (b) double edges (DE), (c) duplicate points (DP), and (d) small cycles (SC). The error correction automatically removes gray edges and white nodes with dashed borders. Black nodes with white borders are merged into edge points, while white nodes with black borders remain nodes. An exception is the red edge in SC4. To correct this error, this edge must be inserted manually.}
\label{fig:errors}
\end{figure*}

\paragraph{False branches (FB)} arise when the skeletonization has nodes/edges that do not represent actual branches in the 3D representation. These are corrected manually as the graph data has no record of the original segmentation. The only way to identify a false branch is by visually detecting that the branch is not part of the 3D representation.  False branches can be separated into 4 categories FB1-FB4 (Fig.~\ref{fig:errors}a).
\begin{itemize}
\item FB1 occurs when branches form a ``v" at the root node. 
\item FB2 occurs when an edge branches off of a straight vessel.
\item FB3 occurs when a vessel further down the tree branches upward close to the first bifurcation, causing a false connection.
\item FB4 occurs when a branch connects the first two daughter vessels, forming a cycle.
\end{itemize}
\paragraph{Double edges (DE)} occur when two nodes are connected by two different edges. The longer of the two edges is removed (Fig.~\ref{fig:errors}b).

\paragraph{Duplicate points (DP)} occur when an edge $E_{(a,b)}=\{[a,b]:e_1,e_2,...,e_m\}$ has $e_i=e_{i+1}$. In this case we delete the duplicate points $e_{i+1}$. This error is shown in Fig.~\ref{fig:errors}c.

\paragraph{Small cycles (SC)} are associated with nodes of degree 4. The merging protocol (Fig.~\ref{fig:merge}) is meant to correct all small cycles, but only works for cycles containing nodes of degree $\leq3$. Therefore, some SC remain after the merging protocol. To correct these, we apply Dijkstra's shortest path algorithm~\cite{Dijkstra59, Kirk07}. Let $a_{root}\in N_G$ be the ID of the root node in a graph $G=(N_G,E_G)$. A \emph{path of length} $p$ is an ordered list of nodes $[a_1, a_2, ..., a_p]\subseteq N_G$ such that any 2 nodes $a_k, a_{k+1}$ are connected via an edge in $E_G$. For each $a\in N_G$, Dijkstra's algorithm returns the shortest path from $a_{root}$ to $a$. We denote this path $P_a$ and the set of all these paths for a graph as $\bm{P_G}$.
\begin{equation}\label{eq:paths}
    \bm{P_G}=\{P_a=[a_1, a_2, ..., a_p] \}
\end{equation}
$$\forall P_a\in \bm{P_G},\quad \begin{cases} a_1=a_{root} \\
a_p=a\quad \end{cases} $$
For all edges $E_{(a,b)}\in E_G$, if no path in $\bm{P_G}$ has the nodes $a$ and $b$ listed consecutively. Shor cycles are broken by removing the edge from $E_G$. 

The small cycles include 4 types SC1-SC4 (Fig.~\ref{fig:errors}d). 
\vspace{-0.2cm}
\begin{itemize}
\item SC1s are found in bifurcations where the bottom node of one of the daughter vessels has degree 4. \\[0,1cm]
\item SC2 occurs when a cycle forms between two daughters in a trifurcation. 
\item SC3 occurs when 2 cycles form between 2 pairs of daughters in a trifurcation. 
\item SC4 occurs at a bifurcation if the bottom nodes of each daughter are the top nodes of edges ending in the same bifurcation node, causing a loop of 4 voxels. 
\end{itemize}
In SC1-SC3, the protocol described above automatically corrects the cycles. For SC4s, it is necessary first to create an edge manually (shown in red in Fig.~\ref{fig:errors}d), and then use Dijkstra's algorithm to break the cycles. 

After error correction is complete, some nodes of degree 2 may exist as a result of removing edges. We convert these to edge points of a newly defined edge connecting the two adjacent nodes (Fig.~\ref{fig:errors}).

\subsection{Network generation}

The result of the graph extraction process is a tree with the same branching pattern as the arterial tree. The edges in this tree are one voxel in diameter, and each edge represents a vessel between junction nodes (or between a junction and a terminal node). The next steps needed to generate an arterial network are: 1) orientating the edges, 2) obtaining the vessel radii, and 3) obtaining the vessel length.

\paragraph{Edge orientation:} Since blood flow in the pulmonary arteries has a direction, we represent the network by a directed graph, i.e., every edge must have a ``start" node and ``end" node indicating the direction of blood flow through that vessel. The graph extraction described above does not distinguish between start and end nodes, i.e., for a given edge $E_{(a,b)}\in E_G$ it is not necessarily true that the edge starts at node $a$ and ends at node $b$. We use the paths $\bm{P_G}$ (equation (\ref{eq:paths})) to obtain proper vessel orientations. If $E_{(a,b)}\in E_G$ with $k=vID\left(E_{(a,b)}\right)$ and $P_b\in\bm{P_G}$ has length $p_b$, then the orientation of $E_{(a,b)}$ is given by
\begin{equation}\label{eq:orientation}
    o(k)=\begin{cases}
1 & \text{if } P_b(p_b-1)=a \\
-1 & \text{otherwise}
\end{cases}
\end{equation}
In other words, $o\left(E_{(a,b)}\right)$ is $1$ if $a$ is the start node and $-1$ if $b$ is the start node. 

We use the orientation to correct the order of the nodes for each $E_{(a,b)}\in E_G$. If $o\left(E_{(a,b)}\right)=1$, we leave $E_{(a,b)}$ unchanged. If $o\left(E_{(a,b)}\right)=-1$, we replace $E_{(a,b)}=\{[a,b]:e_1,...,e_m\}$ with the edge $E_{(b,a)}=\{[b,a]:e_m,...,e_1\}$. In the resulting network, any edges  $E_{(a,b)}\in E_G$ has the start node ID $a$ and the end node ID $b$.

\paragraph{Vessel radius:}  Because each voxel $v\in S$ is centered in $F$, the values $d(v)$ from the distance map (equation (\ref{eq:dmap})) are used to determine the radius of the vessel. Within each vessel, a radius value is assigned to each voxel $v\in S$ to be $r_v=d(v)$. Using these values, we compute the radius of the vessel $E_{(a,b)}$ with $k=vID\left( E_{(a,b)} \right)$ to be $r(k)$ from the inter-quartile mean of the radii measurements $R_k=\{r_v|v\in E_{(a,b)}\}$. We sort $R_k$ from lowest to highest and renumber them as $r_1\leq r_2\leq ...\leq r_{m_k}$ and assign the  radius to each vessel $k$ as
\begin{equation}\label{eq:IQM} r(k)=\frac{\sum_{i=1}^{m_k}w_ir_i}{\sum_{i=1}^{m_k}w_i}
\end{equation}
$$ w_i=\begin{cases}0, & i\in[1,H]\cup[m_k-H+1,m_k] \\ 
1, & i\in[H+2, m_k-H-1] \\
R-4H, & i\in\{H+1,m_k-H\}
\end{cases} $$
$$ R=m_k/4, \quad H=\lfloor R\rfloor$$

\paragraph{Vessel length:} The vessel length $L(k)$ is computed by adding up the Euclidean distances (equation (\ref{eq:euclid})) between each pair of consecutive points along each vessel $E_{(a,b)}=\{[a,b]: e_1,...,e_m \}$ with $k=vID\left( E_{(a,b)} \right)$.
\begin{equation}\label{eq:length}
    L(k)=\left\Vert a-e_1 \right\Vert_2 + \sum_{i=2}^m \left\Vert e_{i-1}-e_i \right\Vert_2 + \left\Vert e_m-b \right\Vert_2
\end{equation}

\subsection{Fluid dynamics domain} 
To conduct fluid dynamics simulations, we characterize vessels as either large or small. Large arteries transport blood to each lobe within the lung, and the small arteries distribute the blood within each lobe.

\paragraph{Principal pathway:} The large arteries form a subtree known as the principal pathway. In the systemic arteries, the ``principal pathway" is formed by named vessels (e.g., the aorta, left and right coronary artery, etc.) distributing blood to all regions in the body. In the pulmonary vasculature, only the right and left pulmonary artery are named, and the vessel morphometry differ significantly between individuals and can therefore not be identified uniquely. However, it is not sufficient to only include the left and right pulmonary arteries as they do not reach each lobe of the lung. To ensure all lobes are reached, we use a scaling argument to select vessels within the principal pathway.

\begin{figure}
\centering
\includegraphics[width=8cm]{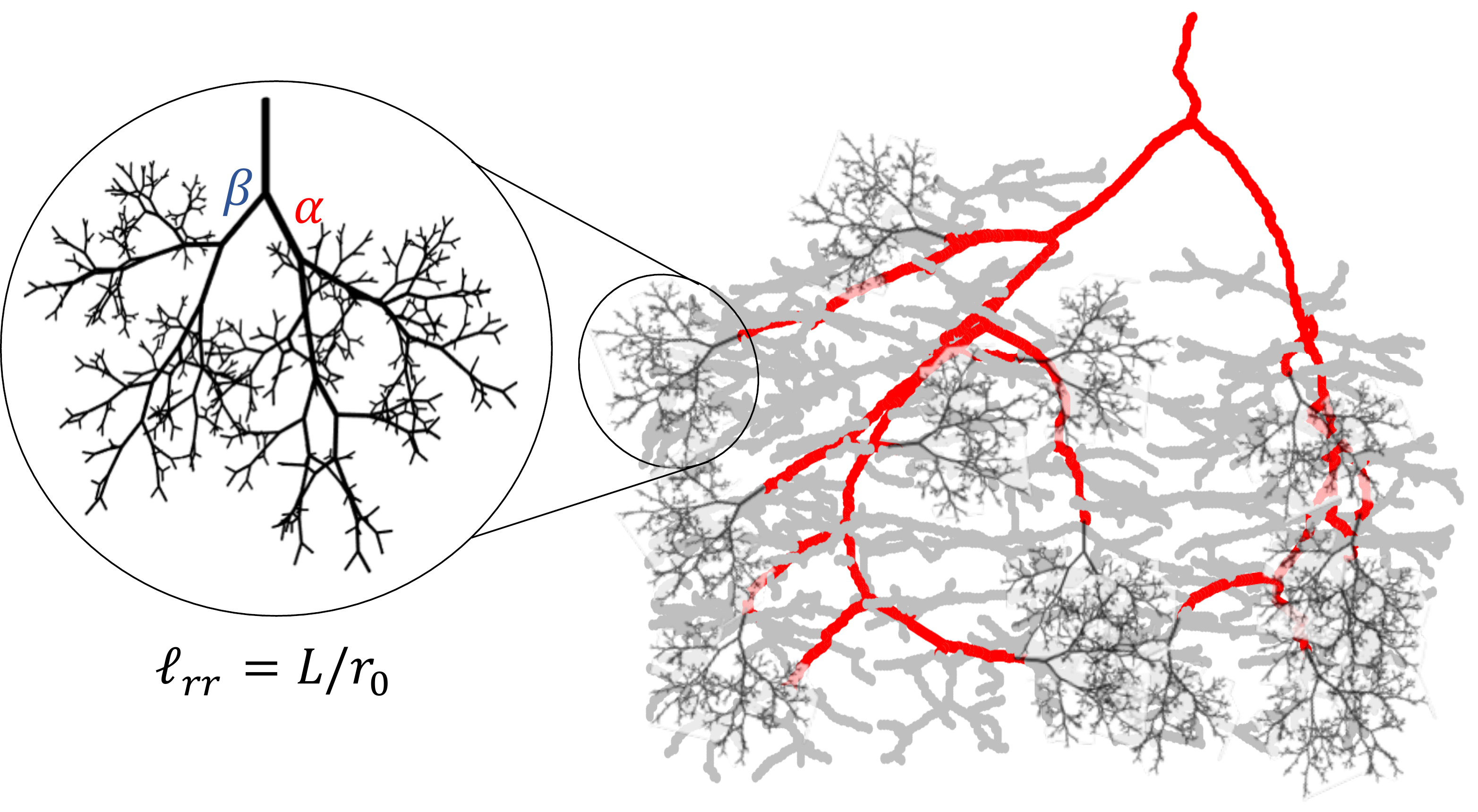}
\caption{Example structured tree. Radii at bifurcations are determined by multiplying the radius of the parent vessel by $\alpha$ (for the daughter of larger radius) and $\beta$. In our fluids simulations, structured trees are attached at all terminal nodes of the principal pathway, shown on the right in red.}
\label{fig:st}
\end{figure}

Let $k_{root}$ be the vessel ID of the root vessel. The principal pathway, denoted by $PP=(N_{PP}, E_{PP})$ is a subtree of $G$. For each edge $E_{(a,b)}\in E_G$, if $k=vID\left(E_{(a,b)}\right)$ and
\begin{equation}\label{eq:pp} \frac{r(k)}{r(k_{root})}\geq 0.4,
\end{equation}
we include $a,b\in N_{PP}$ and $E_{(a,b)}\in E_{PP}$. The largest connected component of this subgraph is retained as the principal pathway $PP$, and any disconnected branches are eliminated (Fig.~\ref{fig:st}).

\paragraph{Structured tree:}
Vessels off the principal pathway are represented as structured trees in which the daughter vessels are scaled by factors $\alpha$ and $\beta$ relative to their parent vessel (Fig.~\ref{fig:st})~\cite{Murray26, Zamir78, Olufsen00}. 

At each bifurcation within the directed network, let $p$ denote the vessel ID of the parent vessel, $d_1$ the ID of the daughter vessel with the larger radius, and $d_2$ the ID of the other daughter vessel, i.e.,
\begin{equation}\label{eq:ab}
	    r(d_1)=\alpha\, r(p),  \quad r(d_2)=\beta\, r(p)
\end{equation}
The length to radius ratio of a vessel $k$, $\ell_{rr}(k)$, is defined by
\begin{equation}
\ell_{rr}(k)=L(k)/r(k) 
\label{eq:lrr}
\end{equation}
To determine $\alpha$ and $\beta$, we analyze the branching structure of all vessels off the principal pathway.

\subsection{Fluid dynamics}
To simulate pulmonary hemodynamics, we use the 1D fluid dynamics model, predicting flow, area, and pressure in the large and small arteries. In the large vessels, comprising the principal pathway, we solve the 1D Navier-Stokes equations, and in the small vessels, represented by structured trees, we solve a linearized 1D model.

\paragraph{Principal pathway:} Similar to our previous studies ~\cite{Olufsen12,Qureshi18,Colebank19}, we predict flow $q(x,t)$ (mL/s), pressure $p(x,t)$ (mmHg), and area $A(x,t)$ (cm$^2$) in each large vessel within the principal pathway, ensuring conservation of mass
\begin{equation}
    \frac{\partial A}{\partial t} + \frac{\partial q}{\partial x} = 0, 
    \label{eq:mass}
\end{equation}
and momentum
\begin{equation}
    \frac{\partial q}{\partial t} + \frac{\partial}{\partial x}\left(\frac{q^2}{A}\right) + \frac{A}{\rho}\frac{\partial p}{\partial x} = -\frac{2\pi \nu r}{\delta} \frac{q}{A},
    \label{eq:momentum}
\end{equation}
where $\rho = 1.057$ (g/mL) is the blood density, $\nu$ (cm$^2$/s) is the kinematic viscosity, and $\delta = 0.03$ (cm) is the boundary layer thickness, approximated by $\sqrt{\nu T / 2\pi}$, where $T=0.11$ (s) is the length of the cardiac cycle. The right hand-side of the momentum equation (\ref{eq:momentum}) is derived under the assumption of a flat velocity profile with a linearly decreasing boundary layer $\delta$. To close the above system of equations, we relate pressure and area~\cite{Olufsen00} as
\begin{equation}\label{eq:state}
    p(x,t) = p_0 + \frac{4}{3}\frac{E h}{r_0}\left(1-\sqrt{\frac{A_0}{A}}\right)
\end{equation}
where $p_0$ (mmHg) is the pressure at which $A = \pi r_0^2$, $E$ (mmHg) is the Young's modulus, and $h$ (cm) is the wall thickness. The vascular stiffness $Eh/r_0$ is related to the unstressed vessel radius $r_0$, and can be approximated using the functional form 
\begin{equation}\label{eq:stiffness}
    \frac{Eh}{r_0}= k_1 e^{k_2 r_0} + k_3
\end{equation}
where $k_1$ and $k_3$ (mmHg), and $k_2\leq 0$ (cm$^{-1}$) are positive constants. 

Since the system of equations is hyperbolic, each vessel requires two boundary conditions, one at each end of the vessel. Hence, when combining vessels in a bifurcating network, three types of boundary conditions are needed.: 1) at the inlet to the root vessel, 2) at each junction node, and 3) at each terminal node. At the inlet of the root vessel, we prescribe flow informed by data. Hence, at the junctions between vessels, three conditions are needed: a condition at the outlets of the parent vessel and two conditions at the inlet to the daughter vessels. These are obtained by enforcing continuity of total pressure and conservation of flow, i.e.,
\begin{eqnarray}\label{eq:junction}
    p_p(L_p,t) = p_{d_1}(0,t) = p_{d_1}(0,t), \\
   q_p(L_p,t) = q_{d_1}(0,t) + q_{d_2}(0,t)
\end{eqnarray}
which holds $\forall t \in [0,T]$. Finally, the end of each terminal vessel is linked to the structured trees by matching the impedance by relating pressure and flow via the discrete approximation of the convolution integral given by
\begin{equation}
    p(L,t_i) = \Delta t \sum^{M+1}_{k=0} z(0,t_{k})q(L,t_{i-k})
\end{equation}
where $z(0,t_k)$ is the impedance of the structured tree, $\Delta t$ is the magnitude of the time step and $M = \Delta t/T$.

The large artery equations are solved numerically using the two-step Lax-Wendroff finite difference scheme~\cite{Olufsen00}.

\paragraph{Structured tree:} Fluid dynamics in the large arteries of the pulmonary circulation are predominantly inertia driven, whereas viscous forces predominantly influence hemodynamics in the small arteries. The small artery equations do not depend on the nonlinear convective terms, and we assume that the Reynolds numbers in these vessels are sufficiently small so that a fully developed flow is present along the length of the small arteries. Also, we assume that flow and pressure are periodic, i.e.
\begin{eqnarray}
    p(x,t) = \sum^{\infty}_{k=-\infty}P(x,\omega_k)e^{i\omega_k t}, \nonumber \\ \hspace{2mm} q(x,t) = \sum^{\infty}_{k=-\infty}Q(x,\omega_k)e^{i\omega_k t}
\end{eqnarray}
where $i$ is the complex unit, $\omega_k = 2\pi k/T$ (Hz) is the angular frequency, and
\begin{eqnarray}
    P(x,\omega_k) = \frac{1}{T}\int^{T/2}_{-T/2}p(x,t)e^{i \omega_k t}dt,\nonumber \\ \hspace{2mm}Q(x,\omega_k) = \frac{1}{T}\int^{T/2}_{-T/2}q(x,t)e^{i \omega_k t}dt.
\end{eqnarray}

Following the derivations by Olufsen et al.~\cite{Olufsen00,Olufsen12}, we can linearize equations (\ref{eq:mass}) and (\ref{eq:momentum}) in the frequency domain giving
\begin{equation}
    i \omega_k C P+ \frac{\partial Q}{\partial x} = 0,
    \label{eq:mass_freq}
\end{equation}
\begin{equation}
    i\omega_k Q + \frac{A_0 \left(1-F_J \right)}{\rho}\frac{\partial P}{\partial x} = 0.
    \label{eq:momentum_freq}
\end{equation}
Assuming that $Eh\gg pr_0$, the compliance $C$ (cm$^2$/mmHg) can be approximated as
\begin{equation}
    C = \frac{dA}{dp} = \frac{3A_0 r_0}{2Eh}\left(1-\frac{3pr_0}{4Eh}\right)^{-3} \approx \frac{3A_0 r_0}{2Eh}
\end{equation}
where $F_J$ denotes the quotient of first and zero order Bessel functions of the first kind, i.e.,
\begin{equation}
    F_J(w) = \frac{2J_1(w_0)}{w_0 J_0(w_0)}, \hspace{2mm} w_0 = i^3 \frac{r_0^2 \omega_k}{\nu}
\end{equation}
where $w_0$ is the Womersley number. Taking a derivative of equation (\ref{eq:mass_freq}) with respect to $x$ gives the wave equation
\begin{equation} \label{eq:wave}
    \frac{\omega_k^2}{c}Q + \frac{\partial^2 Q}{\partial x^2} = 0, c = \sqrt{\frac{A_0 \left(1-F_J\right)}{\rho C}}
\end{equation}
where $c$ (cm/s) is the pulse wave propagation velocity. Solving equation (\ref{eq:wave}) gives
\begin{equation} 
    Q(x,\omega_k) = a \cos(\omega_k x/c) + b \sin(\omega_k x/c)
    \label{eq:Q_trig}
\end{equation}
where $a$ and $b$ are arbitrary integration constants. Using equation (\ref{eq:Q_trig}) in (\ref{eq:momentum_freq}) gives
\begin{equation}
    P(x,\omega_k) = ig^{-1}\left(-a \sin(\omega_k x/c) + b \cos(\omega_k x/c)\right)
    \label{eq:P_trig}
\end{equation}
where $g = cC = \sqrt{CA_0\left(1-F_J\right)/\rho}$.

Structured tree equations (\ref{eq:Q_trig}) and (\ref{eq:P_trig}) can be solved analytically, taking advantage of their periodic nature. From these equations, we predict vascular impedance in the frequency domain as
\begin{equation}
    Z(x,\omega) = \frac{P(x,\omega)}{Q(x,\omega)} = \frac{i\left( b \cos(\omega x/c) -a \sin(\omega x/c) \right)}{g \left(a \cos(\omega x/c) + b \sin(\omega x/c)\right)},
\end{equation}
i.e., the impedance at the start and end of each small artery is given by
\begin{eqnarray}
    Z(0,\omega) = \frac{i}{g}\frac{b}{a}, \nonumber \\
    Z(L,\omega) = \frac{i\left( b \cos(\omega L/c) -a \sin(\omega L/c) \right)}{g \left(a \cos(\omega L/c) + b \sin(\omega L/c)\right)}.
\end{eqnarray}

Combining these equations allows us to predict the impedance at $x=0$ as a function of $Z(L,\omega)$
\begin{equation} \label{eq:Z0}
    Z(0,\omega) = \frac{ig^{-1}\sin(\omega L/c)+Z(L,\omega)\cos(\omega L/c)}{\cos(\omega L/c) + igZ(L,\omega)\sin(\omega L/c)}.
\end{equation}

The input impedance at the zero frequency, i.e., analogous to the DC component ,  is given by
\begin{equation}
    \lim_{\omega\rightarrow0}Z(0,\omega) = \frac{8\mu L}{\pi r_0^4} + Z(L,0) = \frac{8\mu \ell_{rr}}{\pi r_0^3} + Z(L,0). 
\end{equation}

As the radius of the small arteries decreases, the effects of blood viscosity become more important. Following Pries et al.~\cite{Pries92}, we assume that the viscosity in the small arteries follows
\begin{equation} \label{eq:visc}
    \mu^*(r_0) = \left[1+(\mu_{0.45} - 1)\left(\frac{2r_0}{2r_0 - 1.1}\right)^2\right]\left(\frac{2r_0}{2r_0 - 1.1}\right)^2 
\end{equation}
\begin{equation}
    \mu_{0.45}(r_0) = 6e^{-0.17r_0}+3.2 - 2.44e^{-0.12r_0^{0.645}}
\end{equation}
where $\mu_{0.45}(r_0)$ is the relative viscosity at an average hematocrit level of $0.45$. The above equation was used to match viscosity values in humans, where the viscosity of the large vessels is $3.2$  (g /cm s). To adapt this model to the mouse viscosity of $0.049$, we scale (\ref{eq:visc}) giving $\mu_{ST} = \mu \cdot \mu^*(r_0) / 3.2$.

Similar to the large vessels, at each junction we enforce conservation of pressure and flow giving
\begin{equation}\label{eq:cons_Z}
    Z_p(L,\omega)^{-1} =Z_{d_1}(0,\omega)^{-1}+Z_{d_2}(0,\omega)^{-1}.
\end{equation}

The structured tree is generated using the $\alpha$, $\beta$, and $\ell_{rr}$ values obtained from the analysis of the data. The structured tree bifurcates until the radius of any vessel is less than a specified critical minimal radius value, $r_{min}=0.001$ (cm), where the size of the red blood cell is no longer negligible compared to the size of the vessel. At this radius, we assume that $Z_{trm} = 0$~\cite{Olufsen00}. The structured tree equations are solved by recursively computing the impedance of each daughter vessel starting at the terminal branches. Due to the self-similarity of the tree, we do not recompute the impedance in vessels scaled by $\alpha^i \beta^j$ if it has already been computed previously, speeding up computation.

Finally, to compute pressure along vessels within the structured tree, we use the root impedance in a forward algorithm to predict pressure along the $i^{th}$ branch ~\cite{Olufsen12} as
\begin{equation}
    P_L^i =P_0^i\cos\left(\frac{\omega_kL^i}{c^i}\right)-\frac{iq_c}{\lambda^i p_c} Q_0^i\sin\left(\frac{\omega_kL^i}{c^i}\right),
\end{equation}
where $i=\alpha,\beta, \alpha\beta, ..., \alpha^n\beta^m$, and $n$ and $m$ are the maximal $\alpha$ and $\beta$ branches obtained before reaching $r_{min}$.

\section{Results}

\begin{table}
\small\sf\centering
\caption{Network information}
\begin{tabular}{ m{2.125cm} | m{2.5cm} |m{2.5cm} } 
\toprule 
 & \textbf{Control} & \textbf{Hypertensive}\\
 \midrule
$\#$ of vessels    & $2573 \pm 517$     & $3239 \pm 1103$ \\[0.1cm]
$\#$ of pp vessels & $5.63 \pm 1.34\%$  & $4.49 \pm 2.03\%$ \\[0.1cm]
$\bm{\alpha}$      & $0.883\pm 0.122$   & $0.864\pm 0.135$ \\[0.1cm]
$\bm{\beta}$       & $0.666\pm 0.141$   & $0.655\pm 0.137$ \\[0.1cm]
$\bm{\ell_{rr}}$ ($C_1$) & $13.4\pm 1.2$ & $10.9 \pm 1.1$ \\[0.1cm]
$\bm{\ell_{rr}}$ ($C_2$) & $0.00771\pm 0.00083$ & $0.00797\pm 0.00010$\\[0.1cm]
\bottomrule
\end{tabular}
\label{network_info}\end{table}

\subsection{Structured tree parameters}

\begin{figure}[b!]
\centering
\includegraphics[width=8cm]{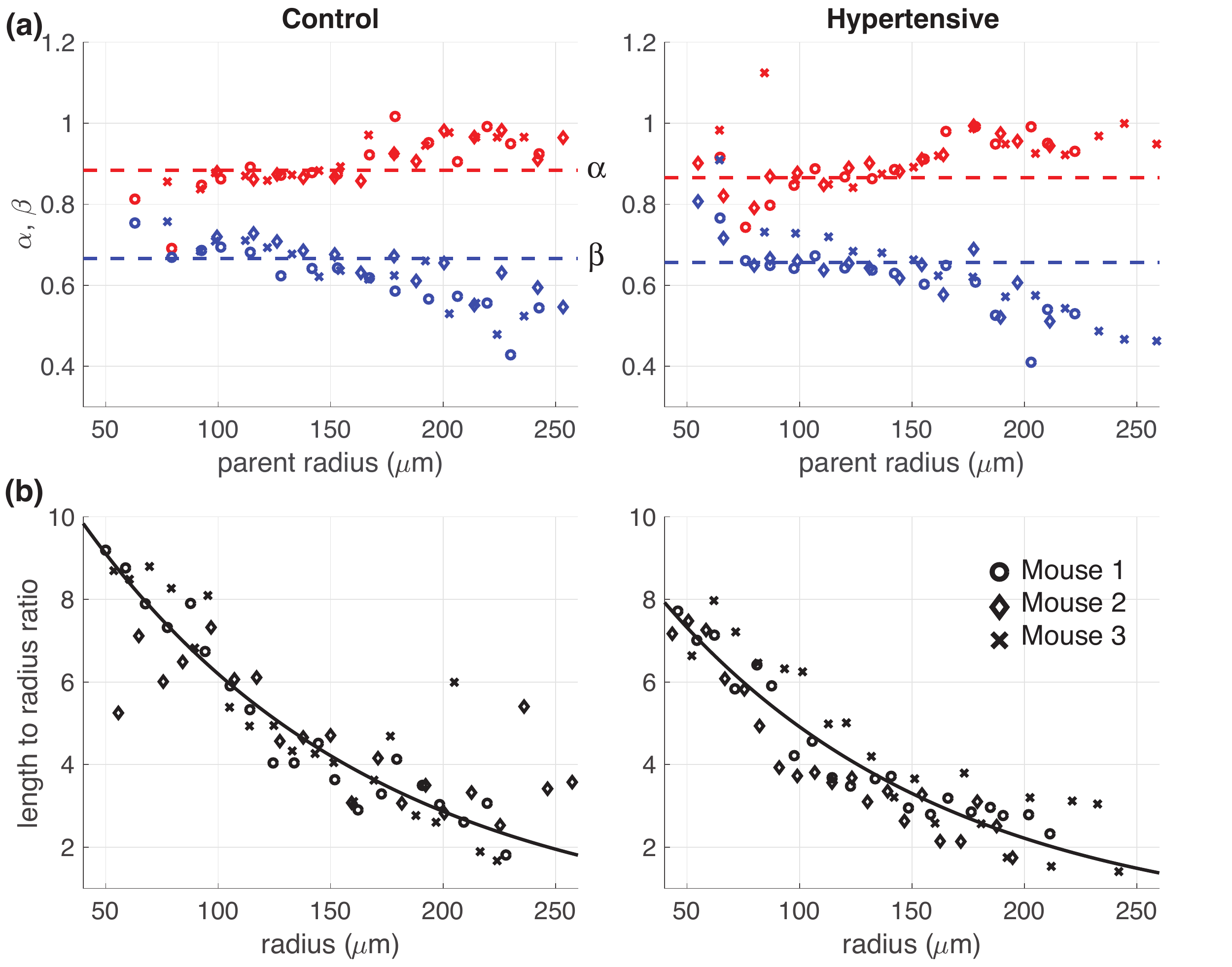}
\caption{Structured tree parameters plotted as a function of the vessel radius. In all graphs, the predictions are grouped into 20 bins, and results include estimates from the 3 control and 3 hypertensive animals. (a) depicts $\alpha$ (red) and $\beta$ (blue), the horizontal lines denote the mean value for each group. (b) depicts $\ell_{rr}$ along with a decreasing exponential curve.}
\label{fig:st_plots}
\end{figure}

Structured tree parameters $\alpha$, $\beta$, and $\ell_{rr}$ are determined in all vessels off the principal pathway (the small vessels). Table \ref{network_info} reports average values over all small vessels in the networks. Figure~\ref{fig:st_plots} depicts parameter values from the data plotted against vessel radius. For each parameter, the observations are divided into 20 bins, and bin averages are plotted at the midpoint of the bin. Both graphs include computations in each of the 3 control and hypertensive animals.  

Fig.~\ref{fig:st_plots}a shows $\alpha$ and $\beta$ as a function of the parent vessel radius. Reference lines are plotted denoting the average $\alpha$ and $\beta$ values for each group of mice. Results show that neither $\alpha$ nor $\beta$ differ between the control and hypertensive animals, and that the subject-specific values agree with values reported by Olufsen~\cite{Olufsen00} ($\alpha=0.9$, $\beta=0.6$).   

Fig.~\ref{fig:st_plots}b depicts $\ell_{rr}$ as a function of the vessel radius. Using a nonlinear least-squares fit with bisquare robustness, we fitted the decaying exponential curve including all points less than three standard deviations from the mean
\begin{equation}\label{eq:exp}
    f(r)=C_1e^{-C_2 r}
\end{equation}
Results show that the length to radius ratio $\ell_{rr}$ decreases with an increasing radius in both control and hypertensive animals.  Values for $C_1$ and $C_2$ are given in Table \ref{network_info}. The predicted values are lower in hypertensive animals compared to controls. This agrees with physiological observations that the radius expands in hypertensive animals, while the length is not affected, resulting in a lower $\ell_{rr}$.

\subsection{Fluid dynamic predictions}
Along the principal pathway, hemodynamic quantities, including blood pressure and flow, are computed by solving the nonlinear fluid dynamics model. For each animal, a measured flow profile is specified at the inlet, and pressure predictions, obtained from solving the fluid dynamics model in the main pulmonary artery, are compared to data from three control and three hypertensive mice. Simulations are done using the subject-specific principal pathway and structured tree parameters $(\alpha,\beta,l_{rr})$.  The  vessel stiffness was applied using values from previously published studies~\cite{Qureshi18,Colebank19}. However, since these earlier studies used a Windkessel boundary condition, parameter tuning was needed to adapt the model to obtain predictions using the Windkessel boundary condition model.  Tuning was done adhering to the constraint that control animals had significantly more compliant vessels than the hypertensive animals, and that stiffness increased in smaller vessels; i.e., the stiffness in the structured trees were larger than in the large vessels. Stiffness values for the principal pathway and structured tree can be found in Table \ref{Tab:Stiffness}.

\begin{table*}
\small\sf\centering
\caption{Stiffness parameters used in model simulations. $PP$-Principal pathway; $ST$ Strutured tree\\}
\begin{tabular}{ m{2cm} | m{2cm} |m{2cm} } 
\toprule 
 & \textbf{Control} & \textbf{Hypertensive}\\
 \midrule
$k^{PP}_1$ (mmHg)  & $129$    & $515$ \\[0.1cm]
$k^{PP}_2$ (cm$^{-1}$)  & $-50$   & $-100$ \\[0.1cm]
$k^{PP}_3$ (mmHg) & $464$   & $902$ \\[0.1cm]
$k^{ST}_1$  (mmHg)& $258$   & $3.86 \times 10^5$ \\[0.1cm]
$k^{ST}_2$ (cm$^{-1}$) & $-60$   & $-50$ \\[0.1cm]
$k^{ST}_3$  (mmHg)& $412$ & $1.29\times10^6$\\[0.1cm]
\bottomrule
\end{tabular}
\label{Tab:Stiffness}\end{table*}

\begin{figure}[b!]
\centering
\includegraphics[width=8.5cm]{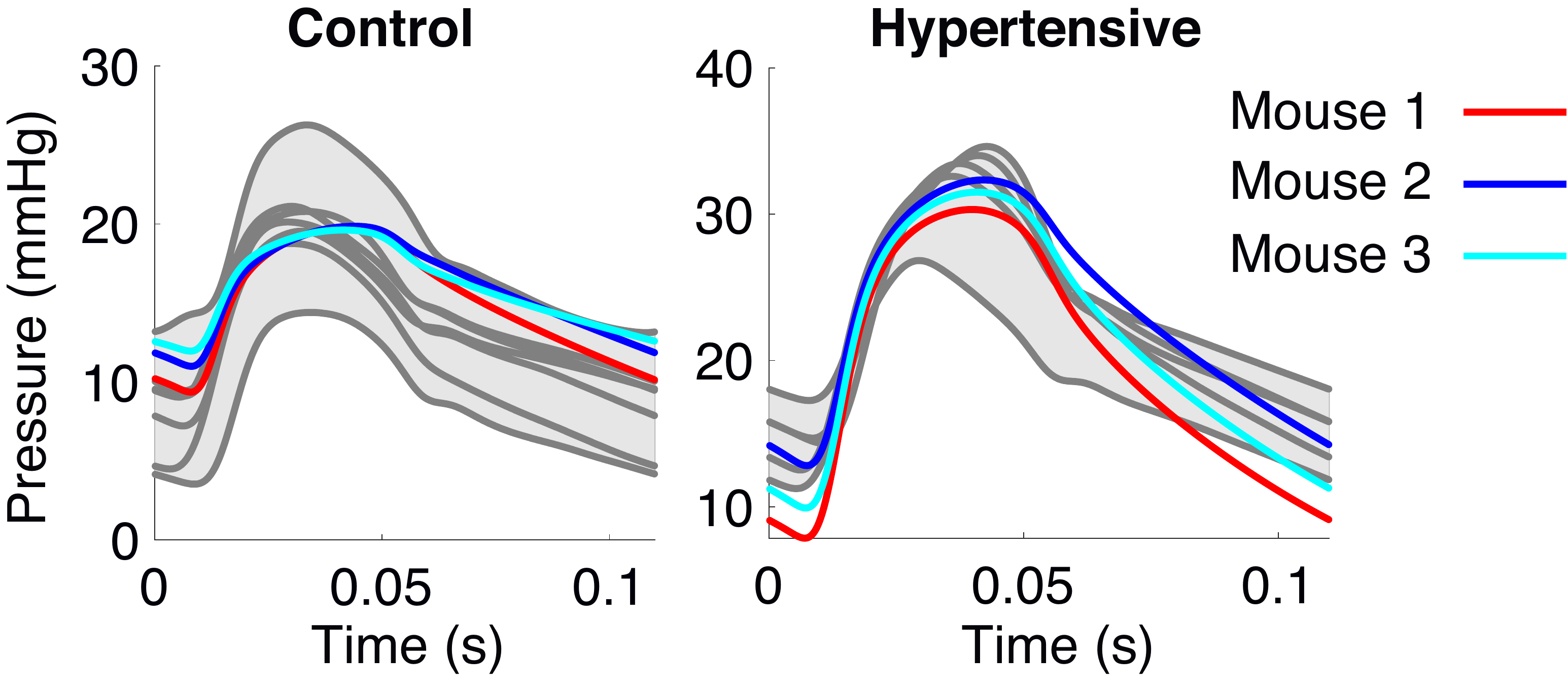}
\caption{Pressure predictions in the control (left) and hypertensive (right) mice compared to measured pressure wave forms for the three mouse geometries.}
\label{fig:pressure}
\end{figure}

\begin{figure*}
\centering
\includegraphics[width=\textwidth]{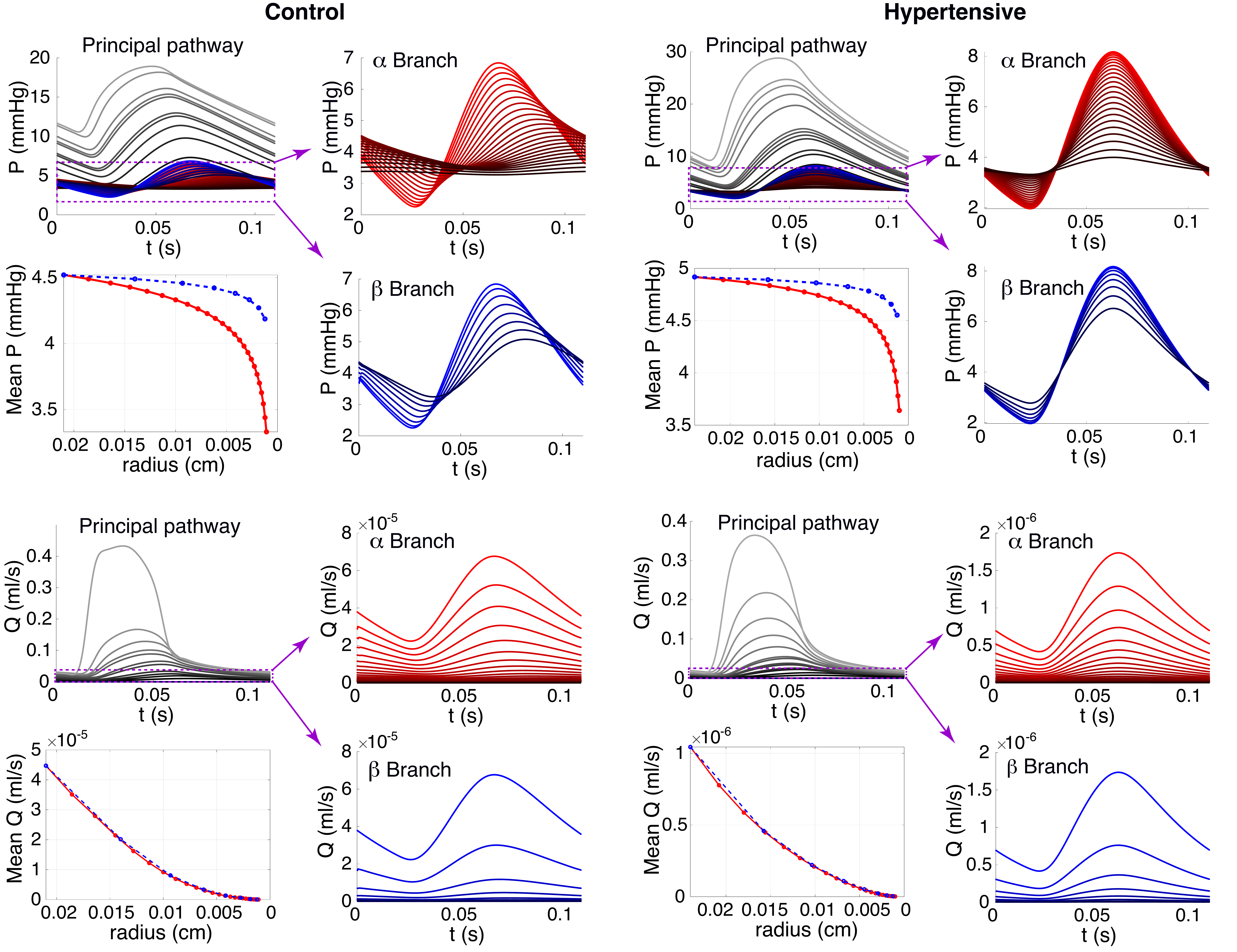}
\caption{Pressure and flow predictions along the principal pathway and $\alpha$/$\beta$ structured tree branches in a control and an hypertensive mouse. Mean pressure and flow predictions in the structured tree model are also provided. Note that the pressure drop along the principal pathway is more sudden compared to the pressure drop in the structured tree. Principal pathway predictions run from the MPA (lightest grey) to the terminal vessel (darkest grey) proximal to the structured tree. Structured tree predictions run through the $\alpha$-branch (marked in red) and the $\beta$-branch (marked in blue), with higher generations denoted by lower magnitude pressure and flow values, as well as a darker color.}
\label{fig:ST_pressureflow}
\end{figure*}

Figure~\ref{fig:pressure} shows pressure predictions in the MPA for each network along with measurements from 5 control and 7 hypertensive animals. Results show that model predictions agree well with data and that by changing vessel stiffness, we can predict pressure in both control and hypertension. Figure~\ref{fig:ST_pressureflow} shows changes in the pressure and flow profiles along the principal pathway as well as predictions within the $\alpha$ and $\beta$ branches of the structured tree for a control and hypertensive mouse. Predictions in the $\alpha$ and $\beta$ branches correspond to vessels with radius $\alpha r_0, \alpha^2 r_0, ..., \alpha^n r_0$ on the $\alpha$ side and $\beta r_0, \beta^2 r_0, ..., \beta^m r_0$ on the $\beta$ side, where $n$ and $m$ are the number of branches before reaching a radius less than $r_{min}$. The mean pressure and mean flow drop along the $\alpha$ and $\beta$ branches within the structured tree are also shown in Fig.~\ref{fig:ST_pressureflow} as a function of radius. Closer scrutiny of the pulse pressure predictions in the MPA (Fig.~\ref{fig:pressure}) show that the model predictions provide a better agreement with data from hypertensive animals than for the controls. In particular, control predictions overestimate diastolic pressure values, while the hypertensive model  underestimate pressure predictions at diastole. The mean pressure drop in both disease states decreases nonlinearly as a function of the structured tree radius and drops quicker in the $\alpha$ branch than the $\beta$ branch. 

\begin{figure*}
\centering
\includegraphics[width=\textwidth]{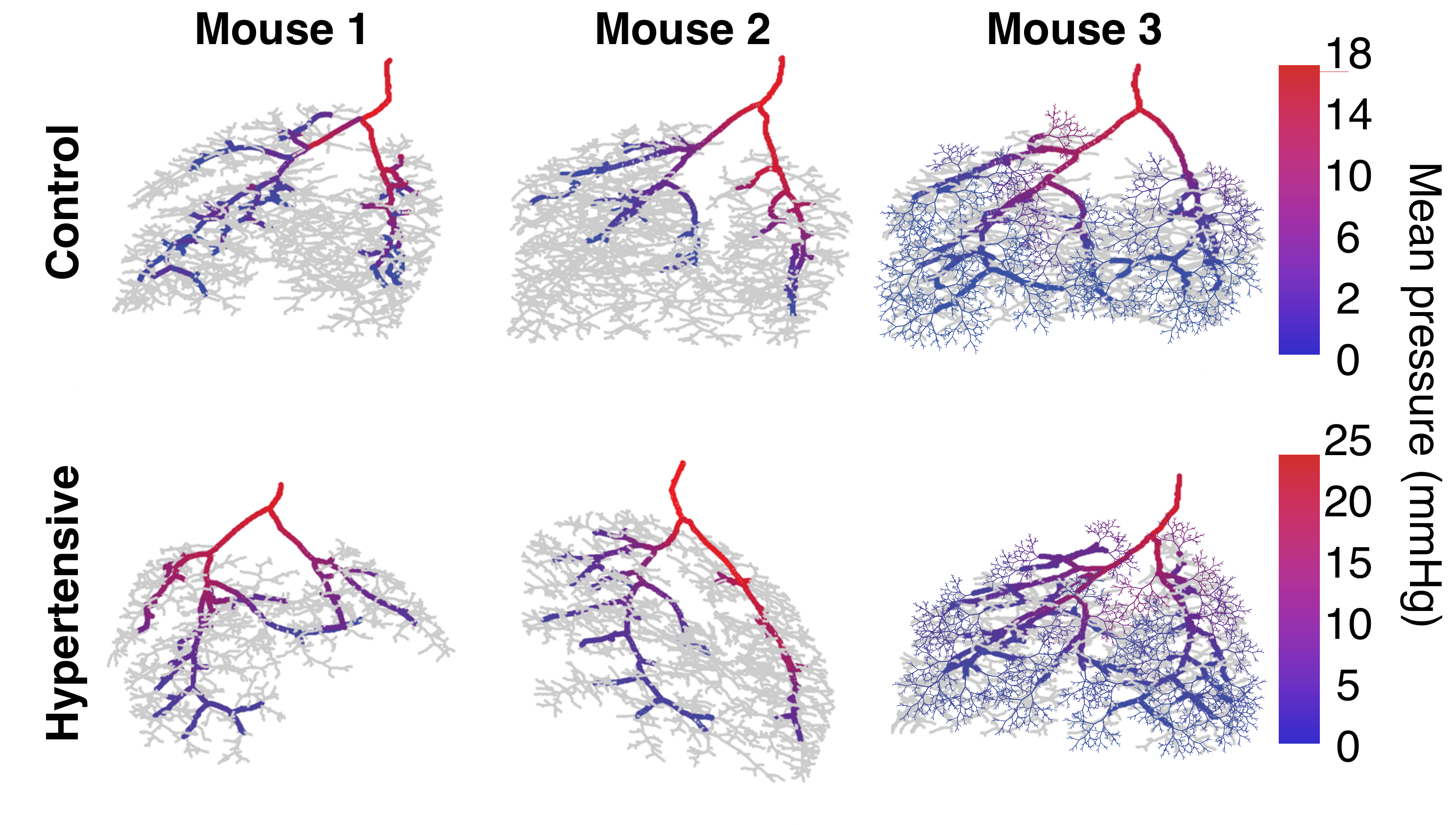}
\caption{Pressure predictions in the three control and hypertensive mice. For mouse 3 we show predictions both along the principal pathway and the structured trees, whereas for mice 1 and 2 we only show predictions along the principal pathway. To illustrate results the structured trees were rendered in 3D, though computations are all done in 1D as described in the method section. It should be noted that the principal pathways perfuse every lobe of the lung, but since the 3D image is projected in 2D parts of the principal pathway lie behind the small vessels marked in gray.}
\label{fig:pp}
\end{figure*}

A benefit of the 1D model is that it is easy to predict pressure in both large and small vessels. Figure~\ref{fig:pp} depicts the mean pressure along the principal pathway and structured trees in the 3D network. In  the control animals, the mean pressure drops from approximately 17 mmHg in the MPA to 1 mmHg at the level of the capillaries, while in the hypertensive animals, the mean pressure changes from 24 mmHg to 2 mmHg.

\section{Discussion}

\subsection{Network reconstruction}
We have presented a network extraction process that captures the length, radius, and connectivity of vascular networks using an accurate, semi-automated method. Our investigation into exception handling for skeletonized arterial trees is, to our knowledge, the first of its kind. Several previous imaging studies~\cite{Burrowes05,Helmberger14,Colebank19} have reported that segmentation leads to false loops and extra branches, but have not been successfully addressed how to remove these from the generated graphs. A previous study by Miyawaki et al.~\cite{Miyawaki17} addressed how to handle trifurcations in bronchial networks, but did not address how to remove loops and false branches, an issue that likely would arise if more generations were included in the network.

One of the most significant benefits of our method is its ability to determine the inlets and outlets of the network automatically. While several pulmonary modeling studies~\cite{Spilker07,Qureshi18,Colebank19} have had success in using open source centerline algorithms, such as the Vascular Modeling ToolKit~\cite{Antiga08}, the studies required manual identification of the terminal vessels of the tree. This quickly becomes infeasible for large networks, such as the pulmonary tree, as there exist hundreds to thousands of terminating arteries. Our methodology identifies all terminal vessels in the tree automatically, making network reconstruction more efficient. This technique can be applied to human-based pulmonary studies, but could also be applied to other biological networks that have a rapid branching pattern, such as the liver~\cite{Ma19} or the brain.

Another use of the results reported here is to provide a validation of geometry that can be used in studies generating artificial networks, e.g., as was done by Clark et al.~\cite{Clark18}, who used a CT image to generate large vessels combined with a space-filling algorithm to generate small vessels. 

\subsection{Structured trees}
The results presented here provide a first to attempt to validate the structured tree model, first introduced by Olufsen~\cite{Olufsen99}. While several previous studies have implemented the structured tree in the pulmonary circulation~\cite{Clipp09,Olufsen12}, the algorithm has never been validated for subject-specific networks.  Moreover, previous studies utilizing the structured tree model~\cite{Clipp09,Olufsen12,Qureshi14} were formulated using three fractal parameters: the radius relation (also known as Murray's exponent) $\xi$, the asymmetry ratio $\gamma$, and the area ratio $\eta$. These values are used in concert to construct estimates of $\alpha$ and $\beta$ using literature values aggregated from multiple patients. Our methods extract $\alpha$ and $\beta$ directly from the data,  minimizing the possible variability that is introduced by coupling network data and literature-based fractal properties. Some key findings in this study are that the parent-daughter radius scaling factors ($\alpha$ and $\beta$) remain relatively constant throughout the lung, and that the subject-specific values agree with values reported by Olufsen~\cite{Olufsen00} ($\alpha=0.9$, $\beta=0.6$).

In this study, we assume that $\alpha$ and $\beta$ are constant, but it can be argued that the lung has two zones separated by $r\approx150 \mu$m. This result agrees with suggestions by Clipp et al.~\cite{Clipp09} analyzing pulmonary arterial casts in a lamb. This study predicted $\alpha$ and $\beta$, but did estimate the $\ell_{rr}$. Instead, they used optimization to best match measured pressure waveforms with model predictions from a 1D fluids model. 

A clear benefit of our method is that we do not require assumptions of fractal parameters based on literature values, making it easy to extract subject-specific, region-specific, and disease-specific values. Results reported here did not identify differences between lobes or with disease. This may be a result of genetic similarity among laboratory animals, or it could be proof that the morphometry of the vasculature favors specific optimality principles. 

Another accomplishment of our algorithm is detection of the principal pathway. In the systemic arteries, the major arteries (e.g., the major aortic branches) are named, making it easy to determine what vessels to include in a modeling study. The same is not valid for the pulmonary network. The MPA artery branches into two vessels that transport blood to the left and right lung. At this stage, it is essential to perfuse every lobe of the lung, but as shown in Fig.~\ref{fig:pressure}, significant variation is observed between individuals, making it challenging to identify specific principal pathway vessels. In this study, we used a scaling law, including all connected vessels with radius at least $40\%$ of the root vessel radius in the principal pathway. This method is advantageous over previous methods, e.g., by Molthen et al. ~\cite{Molthen04}, who only identified one pathway in each lung. Moreover, this method, coupled with the calculation of the radius using the interquartile mean, is robust to image segmentation uncertainty, where vessel radii might be larger depending on image segmentation parameters~\cite{Colebank19}.

\subsection{Fluid dynamics}

The results presented here highlight the advantages of a multiscale modeling approach. Using the image generated principal pathway network, we get a subject-specific geometry in which we can solve nonlinear fluid dynamics equations. By combining this geometry with a subject-specific structured tree model, it is possible to retain physiological features of the large arteries while still modeling microcirculation structure and function. This approach generates a sophisticated model at a relatively low computational cost, advancing previous studies (e.g.,~\cite{Clark18}) that used large networks but did not account for inertial effects in the large vessels, or studies by Colebank et al.~\cite{Colebank19,Colebank19a} that used lumped parameter boundary conditions and therefore was unable to predict dynamics in the microcirculation.  The latter is of importance when using computational methods to analyze disease or to study model-based approaches to treatment when disease progression affects the microcirculation.

In contrast to previous pulmonary studies utilizing the structured tree model~\cite{Clipp09,Olufsen12}, our results are the first to couple a large pulmonary tree segmented from imaging data with a structured tree model of the microcirculation that is also conditioned on data. Our unique approach of including the largest vessels in the principal pathway ensures that the model transports blood to all major lobes of the lung without having to solve computationally expensive, nonlinear equations in the entire pulmonary tree. The identification of the principal pathway is critical, in particular in studies aiming at predicting pulmonary perfusion. This fact has only been addressed in a few studies~\cite{Karau01,Molthen04}, but has been known for years. For example, an early study by West et al.~\cite{West69} showed that the hydrostatic pressure differences in the lung are the driving factor in heterogeneous lung perfusion. While we do not account for lobe-specific pressure differences, the model presented is able to predict pressure and flow within different lobes of the lungs, and can be validated with experimental results in future studies. Our predictions in the MPA show that the hypertensive predictions are, on average, closer to the data than the control predictions. By increasing the large and small artery stiffness, the computational model can predict accurate systolic pressure values in hypertension induced by hypoxia. Several previous studies~\cite{Simonneau04,Vanderpool11,Tabima12} have documented that increased small artery stiffness is a biomarker of pulmonary hypertension, and it is suspected to play a significant role in the increase in right ventricular afterload. Hypertensive results in Fig.~\ref{fig:pressure} are created in part by increasing small artery stiffness by several orders of magnitude relative to the control stiffness (shown in Table \ref{Tab:Stiffness}). This increase in small artery stiffness agrees with current physiological knowledge and previous modeling studies of the pulmonary circulation~\cite{Olufsen12,Qureshi14}. The flexibility of adjusting both large and small artery stiffness in the model allows systolic, diastolic, and pulse pressure values to be altered based on downstream tissue characteristics, making the parameters easier to interpret than standard Windkessel boundary conditions~\cite{Qureshi18,Colebank19}. Pressure predictions in both control and hypertensive networks decrease rapidly from the main pulmonary artery down to the terminal arteries, and show a slower decay in the structured tree predictions. While this is uncharacteristic of the systemic circulation~\cite{Olufsen00}, the pulmonary circulation shows a uniform pressure drop across the rapidly branching arterial tree~\cite{Boron17}. The principal pathway flow predictions in Fig.~\ref{fig:ST_pressureflow} show a similar behavior, dropping in magnitude rapidly from the MPA to the termnal arteries. In contrast to the pressure, the flow distribution in the structured tree tends to be consistent in both the $\alpha$ and $\beta$ branches, whereas the mean pressure drop is more predominant in the $\alpha$ branch than the $\beta$. These results are similar to the findings by Olufsen et al.~\cite{Olufsen12}, showing a more pronounced drop in the $alpha$ side of the tree as this side contains more vessels. For computational ease, only the $\alpha$ and $\beta$ sides are provided, yet the structured tree model can predict pressure and flow in any of the branches of the structured tree.

\subsection{Limitations}

There are several limitations of the study and the results presented here. First, we did not consider the effects of hypoxia on blood viscosity and hematocrit. Previous investigations of blood composition during hypoxia revealed that hematocrit levels increase from the typical value of $0.45\%$ to almost $70\%$~\cite{Schreier14}. Including disease-specific viscosity values based on hematocrit could reveal additional factors that cause an increase in pulmonary vascular resistance and increased pressure in the MPA. 

The computational model uses constant subject-specific values of $\alpha$ and $\beta$, but the data presented could be fitted by a sigmoidal or bi-modal function. Finding appropriate functional forms for these two fractal parameters will be pursued in future studies to ensure that the network becomes more symmetric for smaller radius values. Also, the structured tree model only include bifurcations, but some trifurcations are present in the data (type SC2-SC4 in Fig.~\ref{fig:errors}d). However, only $1.67\%\pm 0.61$ of the junctions in the control and $1.86\%\pm 0.59$ in the hypertensive mice were non-bifurcations (mostly trifurcations, with one quadfurcation in one of the hypertensive networks). Therefore, we skipped non-bifurcating junctions when calculating $\alpha$ and $\beta$ in the networks.

We observed more vessels in the hypertensive networks. This is likely a result of the imaging/segmentation process. In these animals, as expected, vessels expand, and therefore more vessels are visible in the images (above the threshold for imaging), one way to prevent this is to limit the segmentation to a specific number of vessels or generations. Moreover, we did not quantify the branching angles necessary to project fluid predictions in 3D, nor did we employ formal parameter estimation or sensitivity analysis techniques, mostly since imaging and hemodynamic data are not from the same animals. However, the techniques presented here can easily be extended to account for these factors, as discussed in some of our previous studies~\cite{Qureshi18,Colebank19,Colebank19a}. Finally, we used a zero impedance terminal boundary condition, which satisfies the experimental protocol for the imaging data, as vessels are not inflated at the arteriole level. However, if adapted to \textit{in-vivo} studies, this condition should be modified, as suggested by Clipp et al.~\cite{Clipp09} who used included the variation with respiration.

\section{Conclusion}

In this study, we developed a semi-automated method for extracting directed graphs from micro-CT images and used this data to extract a principal pathway and fractal scaling parameters from control and hypertensive mice. These parameters were used to inform a 1D fluids model predicting pressure in the large and small vessels. Results show that the fractal scaling parameters $\alpha$ and $\beta$ do not vary significantly between animals or with disease, but that the length-to-radius ratio $\ell_{rr}$ was lower in the hypertensive animals.  Pressure predictions in the principal pathways were within the range of measured values, and results tuning the vessel stiffness reveal, as expected, that the hypertensive animals have stiffer vessels than the control animals. Also, in both groups the small vessels are stiffer than the large vessels. Our study into network extraction and correction gives us a more thorough understanding of the arterial network structure, and our graph extraction method opens the door for future analysis of arterial networks in various species and organ systems.

\section{Acknowledgements}
The study was supported in part by the National Science Foundation via 
awards NSF-DMS 1246991 and 1615820, and the American Heart Association Predoctoral fellowship 19PRE34380459. In addition, we would like to thank Naomi Chesler, University of Wisconsin-Madison for making micro-CT images and hemodynamic waveforms available for this study.

\end{document}